# A Robust “Shrink-and-wrap” Piecewise Construction Transforms 3-Dimensional Mesh Structures into Mechanical Metamaterials

Fnu Braminder,[1,†] Issac Varghese,[1,†] Noah Kim,[1,†] Alyssa Thomas DeCruz,[,2] and Eric A. Josephs[1,2,*]
[1]Mechanical Metamaterials and Kinetic Devices (Vertically Integrated Projects (VIP)) team; Stony Brook University, the State University of New York (SUNY); Stony Brook, NY 11794
[2]Department of Biomedical Engineering; College of Engineering and Applied Sciences (CEAS); Stony Brook University, SUNY; Stony Brook, NY 11794
[†]These authors contributed equally to this work.
*Correspondence: Eric.Josephs@stonybrook.edu (E.A. Josephs).

## Abstract

Mechanical metamaterials are materials that, by virtue of their microstructural architectures, exhibit mechanical properties not often found in nature: for example, auxetic mechanical metamaterials are materials that possess a negative Poisson’s ratio in response to deforming forces. However, outside of relatively simple crystalline, lattice-based, or repeating-pattern architectures composed of mostly identical unit cells, mechanical metamaterials like auxetics are notoriously difficult to design, particularly across all three dimensions. Here we show that, for any structure that can be decomposed into triangular or tetrahedral meshes, those structures can be subjected to a simple transformation that that creates conditions within those triangular or tetrahedral simplexes that, regardless of their individual geometries, introduces a counter-rotating polygon/polyhedron mechanism that both guarantees a negative Poisson ratio and allows locally programmable mechanical properties. We apply this “shrink-and-wrap” construction to generate mechanical metamaterials from increasingly complex 3D structures, first from simple (convex) polyhedrons; then, going to designs with arbitrarily large numbers of vertices and designed using constructive solid geometry (CSG), computational modeling, and 3D scanning; and, ultimately to 3D objects generated from videos captured by mobile phone camera—making potentially any object or geometric structure designed or found in the real world transformable into a mechanical metamaterial. We show that these monolithic auxetic constructions remain readily “3D printable” via additive manufacturing techniques. Using this family of deformable triangular and tetrahedral unit cells as a basis for the design of complex metamaterial properties, the “shrink-and-wrap” construction represents a unification of progress in computer graphics / computer aided design (CAD), AI-enabled digital structure generation, and the engineering of mechanical metamaterial microstructural architectures. We expect this piecewise approach to designing arbitrarily complex auxetic and mechanical metamaterial structures can enable numerous potential applications where programmable internal reconfigurations and/or force redirection are required across and throughout their 3D geometries.

**Introduction:**

Auxetic metamaterials are materials that exhibit a negative Poisson's ratio (contract rather than expand in the traverse direction in response to a deforming mechanical force, and vice versa) as a result of their having a microstructural architecture that specifically allows for precisely controlled internal reconfigurations enabling that unusual property (reviewed recently in [1-3], see for example Figure S1). Auxetic structures are found in materials requiring, for example, large strength-to-mass ratios and potential for high impact resistance, where internally directed forces within the structure can be optimized for these purposes; this class of materials includes the strongest naturally occurring biological composite materials, the teeth of the Common Limpet (*Patella vulgata*) [4], where auxetic microstructures were recently observed. Auxetic mechanical metamaterials also have a number of emerging biomaterial/biomedical, robotics, and consumer applications where bespoke control of internal force distributions and responses to forces are desired [1-3].

While metamaterial structures with a number of simple geometries are known to exhibit negative Poisson's ratio, including bowtie / reentrant structures among others [1-3], these tend to require a crystalline, lattice-based, or consist of repeating patterns of mostly the same unit cell across their entire structures. The design of non-crystalline auxetic metamaterials or metamaterials having arbitrary structures or designs, let alone those with locally programmable mechanical properties throughout their geometries, remains a notoriously complex challenge [2, 5, 6]. Another mechanism for generating for mechanical metamaterials exhibiting auxetic properties uses networks of counter-rotating polygons (for example, Figure S1) and polyhedrons [7, 8] that, again, can be difficult to match to desired geometries and properties, despite significant recent efforts made towards these ends—in particular, with regards to elaborating the conditions for which 2D and 3D structures will exhibit the auxetic property [7, 8], and for demonstrations that AI-generated designs of counter-rotating auxetic metamaterials can be fine-turned for precise force-deflection responses [6, 9-11]. What would represent a significant advance in the development of auxetic mechanical metamaterials would be a marriage of these approaches: in other words, it would be greatly beneficial if there existed a family of metamaterial microcellular structures that (1) simplified both analysis and design intuition of complex mechanical structures, and (2) could be modularly integrated into diverse geometries, on-demand, to transform any two- or three-dimensional structure or form into an (auxetic) mechanical metamaterial to be readily deployed into their myriad potential practical applications (Figure 1A).

Here, we find that, by imposing a simple design constraint – only that that the structures be decomposable into triangular or tetrahedral simplex meshes [12, 13], which is an extremely common practice in applications of computer aided design (CAD), computer graphics, and computer generated imagery (CGI) as well as constructive solid geometry (CSG) and finite element analysis (FEA) [14] – those structures can be readily converted to mechanical metamaterials via a straightforward construction that we call "shrink-and-wrap" (Figures 1A-D and 2A-E) to describe the two key steps of the procedure. In the "shrink-and-wrap" construction, the simplexes making up a complex geometric structure defined by a mesh can be piece-wise and individually transformed such that they seamlessly connect into a seamless (monolithic) counter-rotating metamaterial microarchitectures (Figures 1, 2, S2, and S3). The piecewise nature of this construction (Figures 1E, 2Fi, 2G-J, S2, and S3) serves as a convenient basis for local programmability with bespoke mechanical properties throughout their geometries (Figures 1F, 1H, and 2F), without suffering from snowballing complexities in their design and analysis as the number of elements increase. We show this robust construction, which by virtue of its geometric constraints automatically creates metamaterial structures with a Poisson's ratio of −1 [7, 8] (the theoretical negative limit for an isotropic material), can be broadly applied not only to simple 2D and 3D geometries but to complex 3D structures generated from CSG, 3D scanning, and even 3D objects generated using videos captured by mobile phone [15]: making potentially any object or geometric structure designed or found in the real world transformable into a monolithic mechanical metamaterial, demonstrating its broad potential applicability.

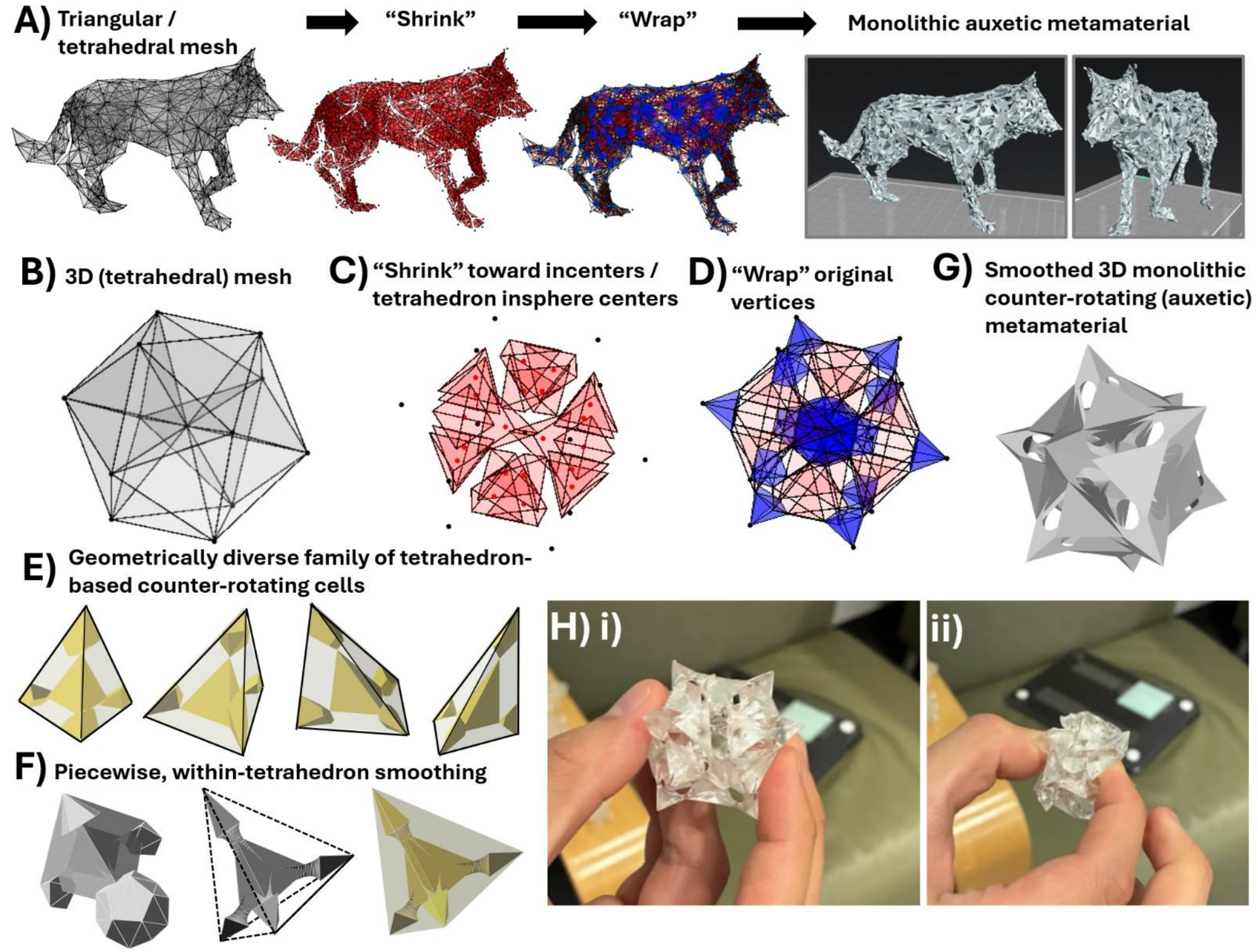

**Figure 1. The "shrink-and-wrap" construction provides a simple way to construct monolithic, counter-rotating (auxetic) mechanical metamaterials out of any geometry that can be decomposed into triangular or tetrahedral meshes.** A) Schematic of the piecewise "shrink-and-wrap" construction of a monolithic auxetic metamaterial structure from any 3D tetrahedral mesh: for example, a wolf tetrahedral mesh generated from a triangular surface found in the stock Microsoft 3D Model Library. B-D) Details of the algorithm to construct 3D auxetics, in analogy with the construction of 2D auxetics shown in Figure 2. B) A set of points (here, a regular convex icosahedron) is used to assemble a tetrahedral mesh. C) "Shrink": the vertices of each tetrahedron are moved toward the tetrahedron's incenter (center of the tetrahedron insphere) to define a new similar tetrahedron each scaled at a factor λ ($0 < \lambda < 1$) relative to the original tetrahedrons, centered on the same incenters. D) "Wrap": a convex hull is constructed around (and inclusive of) each of the original "parent" vertices by connecting the coordinates of the new "child" vertices (now vertices of each "shrunken" tetrahedrons) derived from same parent. E-G) A consequence of this construction is that each original tetrahedron now contains a consistent type of counter-rotating microstructural cell, regardless of original mesh geometry (E): each tetrahedral cell contains a "shrunken" tetrahedron that has, at each vertex, a segment of an overlapping counter-rotating polyhedron "node" (F). By virtue of the construction, the edges and faces of the overlapping segments are co-linear or co-planar across the faces between neighboring metamaterial cells: this allows a seamless, piece-wise assembly of those tetrahedral cells into higher-order, interconnected metamaterial structures. Because of this and the fact that the geometry of the internal microstructures is determined entirely by the points of their tetrahedra and only those points that form touching tetrahedra, shape smoothing can be performed within each tetrahedral cell and individually tuned at each joint. (G). H) A 3D-printed, counter-rotating auxetic mechanical metamaterial object derived from a regular icosahedron (i) prior to applied force, and (ii) collapsing after applying force along one axis, demonstrating its auxetic properties / negative Poisson ratio, i.e., shrinking in the dimensions traverse to the application of force.

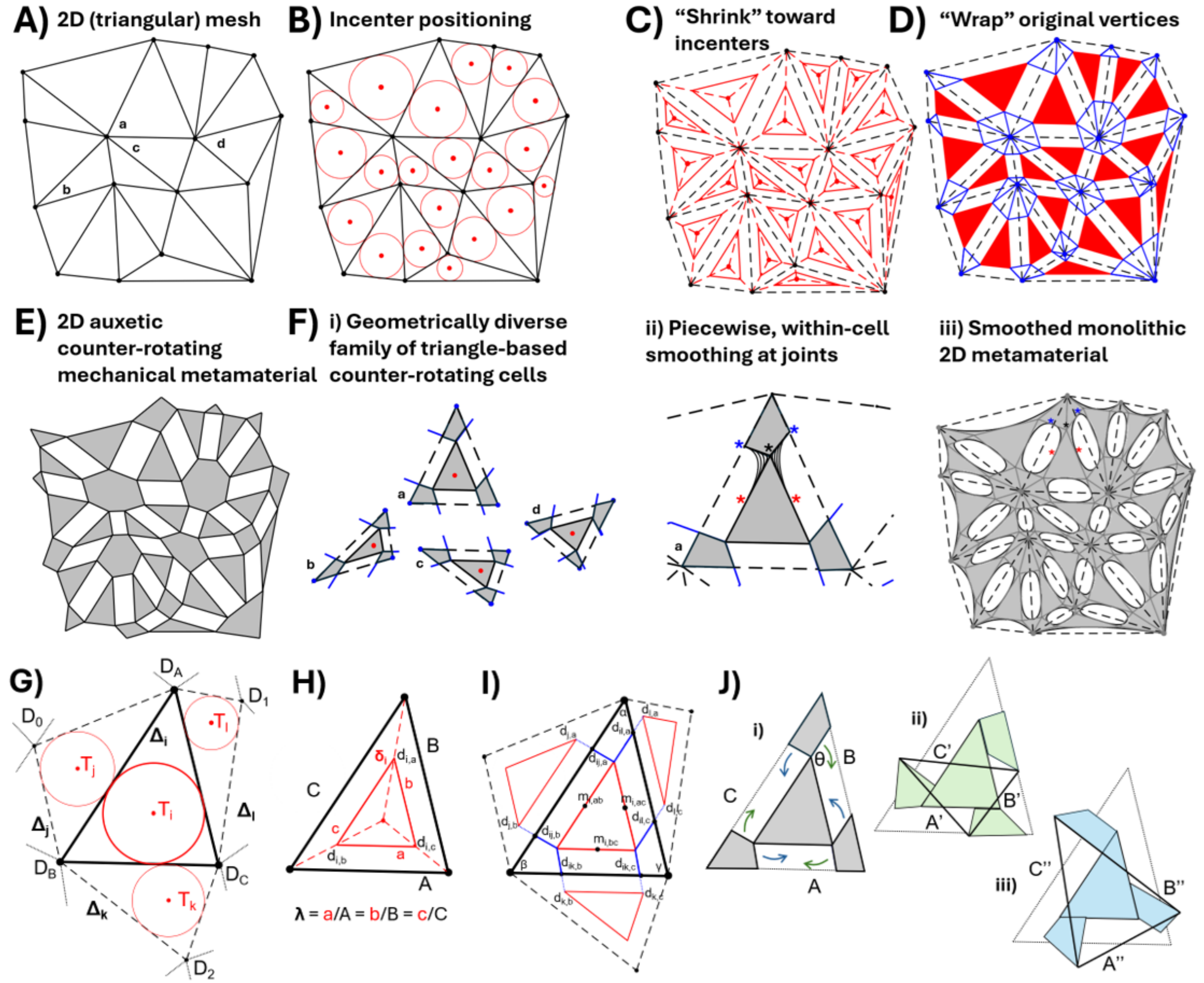

**Figure 2. Illustrative example of the "shrink-and-wrap" construction algorithm in 2D (before extending to 3D).** A-E) Schematic process for the "shrink-and-wrap" construction for 2D- and 2.5D- (2D extruded into 3D space, i.e., see Figure S1). This can serve as a simpler and visualizable basis to expand to the 3D variation, in order to more easily illustrate the basic concepts of the metamaterial construction. Cells marked with a lowercase *a*, *b*, *c*, and *d* are highlighted for Figure 2Fi. (A) Given a (constrained) Delaunay triangulation (or any other triangular mesh), (B) the position of each triangle incenter (center of the incircle) is first located. (C) "Shrink": the "parent" vertices of each triangle are moved along vector between the incenter and that vertex toward the incenter, reducing the length of the vector from the incenter to that new point by a constant factor λ. An interior "shrunken" triangle, similar to the triangle from the original mesh, is formed by connecting the new points. (D) "Wrap": a polygon is formed by connecting the "child" vertices of the shrunken triangles all derived from the same original point, inclusive, "wrapping" that original point. (E) The new set of shrunken triangles and wrapped polygons form the (representative example) auxetic structure derived from the mesh in Figure 2A. F) As in the 3D case, i) regardless of their exact dimensions, the original simplexes (triangles) are each transformed into a member of a consistent family of counter-rotating microstructures (Figure S2)—in this case, having an internal "shrunken" triangle and a quadrilateral at each vertex of the shrunken triangles connecting it to the original vertex. ii) In this construction, it becomes trivial to generate smoothing/fileting at joints between counter-rotating shapes within those cells using Bézier curves, as edges of the overlapping polygons are colinear across the edges of the triangular cells. The red asterisks correspond to the m midpoints on the internal triangle labeled in Figure 2I, the blue corresponding to the $d_{ij}$ and $d_{il}$ points at the edge of the triangular cell, and the black asterisk

corresponding to the “child” vertex serving as the control point ($d_{i,a}$ in 2H). Various smoothing parameters μ (see Methods) with different expected mechanical strengths are shown. iii) Example of a smoothed, monolithic 2D auxetic from 2E. G – J) Geometric description from the vantage of individual metamaterial microstructures for piecewise construction. G) The geometry of each microstructure is entirely determined by the 3 points of the triangle, and the 3 external points forming the 3 neighboring triangles sharing each its faces. H) The shape of the “shrunken” triangle is determined by just the coordinates of the original triangle and ratio λ (Figure S2 and S3). I) The coordinates of 3 quadrilaterals are determined from where the lines connecting “child” vertices of the shrunken triangles with the same “parent” cross each original face. J) Analytically, counter-rotation can understood from within the contexts of individual cells i) at various rotation angles θ. The final structures of each auxetic microstructure after rotation full rotation, ii) either clockwise (green) or iii) counterclockwise(ble) – how far each polygon can rotate relative to the shrunken triangles before jamming (jamming angle) and the final rotated structures – can all be determined at the level of each cell from the coordinates of those six points (Figures S2 and S3).

## Methods

*Code availability.*

Code for this project was written in-house for MATLAB (MathWorks; Natick, MA) and Python and will be shared as a publicly-available distribution upon publication of a peer-reviewed manuscript.

*2D and 3D point geometries, mesh generation, and refinement.*

Point coordinates were generated using scripts for random number generators; identified points describing common convex polyhedrons; publicly available repositories for 3D structures (NIH 3D, 2024. https://3d.nih.gov/; Stanford 3D model repository; Microsoft 3D structure library); generated using CSG with the open-source program OpenSCAD (see notes in [16]); or generated as 3D structures from video captured using an Apple iPhone 16e (Cupertino, CA, USA) with the Gaussian splatting software KIRI engine (Ver 4.2.7; KIRI Engine, Kiri Innovations, Toronto, Canada) app for iPhone. In the case of 2D points, these were often triangulated using constrained or unconstrained Delaunay triangulations [14, 17] using built-in algorithms from MATLAB. In the case of 3D geometries from libraries or generated from KIRI (often triangular mesh skins of 3D structures as STL or OBJ files; 4K resolution), these structures were simplified to ~500 to 1000 surface points (occasionally more as noted) using a Quadratic Edge Collapse Decimation algorithm [18] in MeshLab (2025.07); and an internal geometry of tetrahedrons generated using MATLAB's finite element analysis module for mesh generation or, more often, using the "Fast Tetrahedral Meshing in the Wild" (fTetWild) algorithm [12]. The points and their connectivity in each tetrahedron was then converted for import into MATLAB for auxetic structure generation using Gmsh (4.15.2) [19]. In the case of 3D scanned objects captured using iPhone, any background or support (floor) geometries were removed first manually using MeshLab. After generation of the auxetic metamaterial structures using the "shrink-and-wrap" algorithm (see below) was output to OpenSCAD as a piecewise CSG geometry, was then converted using a Python script to a single complex polyhedron using a in-house written python script for more rapid assembly; often the structures were made "water-tight" using a "Screened Poisson surface reconstruction" [20] routine from MeshLab. These surfaces were then sliced for stereolithographic 3D printing using CHITUBOX (CBD-Tech; Shenzhen Chitu Systems Co., Ltd.).

*Generation of auxetic designs from 2-D and 3-D sets of point geometries using the "shrink-and-wrap" construction.*

The overall design algorithm is described in the main text as well as figures and figure captions for Figures 1 and 2. It is performed piecewise, that is, at the level of each simplex (triangle or tetrahedron) deconstructed for a given triangular or tetrahedral mesh (see above), then ultimately composed into a single monolithic metamaterial structure.

Briefly: for each simplex,

1) the incircle ($T_i$ in the 2D case using the notation in Figure 2G-J) or insphere center position is calculated; for example, in the 2D case [21], in triangle $\Delta_i$ the coordinates for $T_i$ was calculated from the coordinates of points $D_A$, $D_B$, and $D_C$ as $(A*D_A + B*D_B + C*D_C) / (A + B + C)$ where A, B, and C are the side lengths determined from the distances between $D_B$ and $D_C$, $D_A$ and $D_C$, and $D_A$ and $D_B$, respectively.

2) "Shrink": the coordinates of the "shrunken" simplexes are determined by moving coordinate of the "parent" vertex toward the incenter to scale the distance between the incenter and the new "child" by a factor of $\lambda$; in the notation of Figure 2G-J, the coordinates of point $d_{i,a}$, the child of point $D_A$, is calculated as $(1 – \lambda) * T_i + \lambda * D_A$, then repeated for all other vertices. "Shrunken" simplex $\delta_i$ is formed by connecting all the new child points within the original simplex $\Delta_i$.

3) "Wrap": the shape of the polygon (quadrilaterals) or polyhedral segments at each vertex within the original simplexes is then determined. In the 2D case, the coordinates for each quadrilateral at each vertex is determined by the position of the parent vertex (e.g., $D_A$), the child vertex (e.g., $d_{i,a}$), and where the lines connecting the child vertex to child vertices of the two neighboring triangles the faces of the parent vertex cross those two faces (e.g., at $d_{ij,a}$, where the line from $d_{i,a}$ to $d_{j,a}$ crosses the line from $D_A$ to $D_B$, and at $d_{il,a}$, where the line from $d_{i,a}$ to $d_{l,a}$ crosses the line from $D_A$ to $D_c$). Because, in the 3D case, polygons can intersect with tetrahedrons in more complex ways [22], this is simplified here by using calculating the convex hull of the polyhedron formed by all the children vertexes from a parent vertex of a tetrahedron, and performing the CSG Boolean operation for the Intersection of the polygon and the tetrahedron under consideration, which should always include that vertex.

4) “Smooth”: to perform within-cell smoothing / filleting at joints (as in Figures 1F and 2Fii), a smoothing parameter μ (from 0 to 1) is introduced. This allows a precise fine-tuning of joint stiffness, but piece-wise and at the level of individual joints within individual simplexes, in a simplified geometric framework. In the 2D case, for vertex of the shrunken triangle (e.g., at $d_{i,a}$), the midpoints of the two edges sharing that vertex (e.g., $m_{i,ab}$ and $m_{i,ac}$) were calculated parametrically (i.e., $0.5 * (d_{i,a} + d_{i,b})$ and $0.5 * (d_{i,a} + d_{i,c})$, respectively). In the case of full smoothing ($\mu = 1$), a quadratic Bézier curve [23, 24] connecting those midpoints to the neighboring quadrilateral coordinate was constructed with the child vertex as the Bézier control point, i.e., $(1 – t) ^ 2 * m_{i,ab} + 2 * t * (1-t) * d_{i,a} + t ^ 2 * d_{ij,a}$, for values of t from 0 to 1. In the case of μ (<1), a quadratic Bézier curve was constructed using points closer to the child vertex, i.e., $(1 – t) ^ 2 * m'_{i,ab} + 2 * t * (1-t) * d_{i,a} + t ^ 2 * d'_{ij,a}$, where $m'_{i,ab} = \mu * d_{i,a} + (1 - \mu) * m_{i,ab}$ and $d'_{ij,a} = \mu * d_{i,a} + (1 - \mu) * d_{ij,a}$. This construction ensures that smoothing the curves are colinear and continuous along the shrunken triangles at their midpoints and the edges of the original with the polygons in the neighboring triangle. Curves from both edges flanking the child vertexes are combined with the shrunken triangle and associated quadrilateral at that child vertex using the CSG Boolean operator Union to form a continuous geometry. Smoothing in 3D auxetics is performed analogously: for each child vertex, a quadratic Bézier curve is constructed, using the child vertex as the control point, between (i) the triangle incenter of one of the abutting faces on the shrunken tetrahedron and (ii) where the line connecting the child vertex to associated child vertex on the shrunken tetrahedron opposite the associated face on the parent tetrahedron crosses that face. The curves originating from all the faces associated that child vertex are then used to form a Bézier surface that forms the surface of a new polyhedron that is then joined using CSG Boolean operator Union with the internal tetrahedron and the associated polyhedron. Similarly, this enables what would otherwise be complex geometries within these tetrahedrons to be formed at each vertex, piecewise and within each tetrahedral cell.

*3D printing of auxetic designs.*

3D printing was performed using an Elegoo (Shenzhen, China) Mars 5 stereolithographic printer with SuperFlex (ENGR-F12L; 3DMaterials) resin or 70 : 30 mixtures of SuperFlex : Super PP 1.4G (ENGR-P49L; 3DMaterials) resin.

**Results and Discussion.**

It was previously shown [8] that any set of points can be used to construct a counter-rotating network of polygons with auxetic behavior, i.e., that simultaneously collapse (or expand) equally in all directions in respond to a deforming force, as a basis for a mechanical metamaterial with a Poisson's ration of -1, approaching the theoretically most negative value for an isotropic material. In that case, the following conditions were required: (1) the points are partitioned into a bipartite graph where points associated with one group (which we can call **D**, like in Figure 2) exclusively are joined to points of the other group (here **T**); (2) Those points are used to generate polygons with vertices positioned on the lines connecting elements of **D** to elements of **T**; and (3) the position of the vertices along those lines divides the length of the segments from remains constant throughout the network (what we call λ, Figure 2H) [8]. The consequence of this construction is that the gaps/empty spaces between polygons would form parallelograms [7], and the counter-rotating polygons would have the matching angular velocities, allowing maximized rotation before jamming (Figures S3C-F). As derived, this construction was previously only demonstrated in two-dimensional auxetics [8] (or, rather, 2.5-dimensional, with a two-dimensional geometry extruded to a specific depth for the practicalities of 3D printing, as in Figure S1) but should also be applicable for three-dimensional auxetic structures; a recent expansion and extension related to this approach demonstrated these conditions are, in fact, applicable in 3 dimensional structures as well [7]. It was proposed [8] that the bipartite graph could be generated from those original defining points by either (i) starting with a randomly-connected, high-coordination graph, pruning connections between members of the same group and reforming them to members of the other group, until only connections between elements of the different groups were connected. They noted that procedure could be difficult for large and complex network, and result in random or non-deterministic networked structures; or (ii) by directly generating the dual of the graph from the original points, then connecting elements of the dual network exclusively to their generating elements from the original collection. Another recently proposed method involved separating a previously connected network of polygons, then connecting the polygons in a way so the empty space between them was filled with parallelograms, in a manner that also satisfied those above requirements [7].

While those methods to form counter-rotating auxetic networks were quite general, we recognized that a subset of these graphs would exhibit exceptionally well-behaved properties that would allow for precise tuning, analysis, and streamlined application to three-dimensional structures, if we only imposing a simple constraint: that the original structures (formed by **D**) be decomposable into triangular (or tetrahedral) unit cells [12-14] – that is the spaces are divided into (connected) simplexes formed by 3 (2D) or 4 (3D) points such that no other points are present within the circumcircle or circumsphere of each simplex. This constraint – more simply stated as that the structure be defined as a triangular or tetrahedral mesh – is already very common in computer graphics, additive manufacturing, and other computational modeling of 3D structures [14, 17], so it allows direct and ready application to any structures already created for those applications. Additionally, we posited that rather than generate the points **T** (in the opposite members of the bipartite graph) as the dual graph of **D**, instead, they are positioned at the center of each triangular incircle (the center of the interior circle tangent to every edge of the triangle) or center of tetrahedral insphere [21, 25] (Figure S3). Note for the commonly used Delaunay triangulation [14], the dual would instead generate be the points in the graph's Voronoi diagram, placing the dual points in each triangle in the center of each triangle's excircle (its excenter) – rather than incenter; however depending on the precise geometry or dimensions of the simplexes, the excenter can fall outside of the simplexes themselves. The choice of incenter rather than excenter (or triangle centroid, Figures S3a and S3b) therefore forces the counter-rotating polygon or polyhedron (in our case, always triangles or tetrahedrons) to always reside within initial polygon or polyhedron network elements, in turn forming the basis of class of metamaterial microstructures in each simplex regardless of their individual dimensions.

To perform the construction at this stage (starting with a triangular/tetrahedral mesh; Figures 1B and 2A), we call the procedure "shrink-and-wrap" (see Methods; Figures 1A-D and 2A-E) based on the two key steps:

"Shrink" (Figures 1C, 2B-C, and 2G-H: the **D** making up the vertices of each simplex are moved toward their respective incenters to form a similar "shrunken" triangles/tetrahedrons scaled by factor λ (new points **d**) and having the same incenter. This forms the first set of counter-rotating polygons (all necessarily either triangles or tetrahedrons, respectively).

"Wrap" (Figures 1D, 2D, and 2I): the convex hull around all new "child" **d** points derived from the same "parent" **D** point is formed, generating the other set of counter-rotating polygons.

The uniform contraction during the "shrink" stage by λ and the nice geometric properties of the incenter (Figure S3) mean that the geometric conditions for counter-rotating metamaterial properties are automatically met (e.g., Figures 1H) with a theoretical Poisson's ratio of -1 [7, 8], and they also highlight that *the internal structures of each of those new metamaterial cells share a common geometry, regardless of their initial dimensions* (Figures 1E, 2Fi, S2): they all contain a similar scaled simplex with the same incenter as the larger simplex in the determining structure, and that the each vertices of the "shrunken" simplexes all contact the segment of the counter-rotating polygon/polyhedron that overlaps that simplex at each of their corresponding larger vertexes (Figure 1H, 2J, S3, S4). That the individual triangular/tetrahedral elements derived from the larger mesh can be transformed individually into a common set of mechanical metamaterial microstructures reveals a number of exceptional properties, which we enumerate 6 of below:

(1) *Locally-defined metamaterial microstructures:* the geometry constructed within each simplex is determined precisely by just the coordinates the simplex vertices, scaling factor λ, and only the nearest (touching) points: in the 2D case, just the three neighboring points that form triangles sharing a face (Figure 2G-I); and in the 3D case, the structure is determined by the points from neighboring/touching tetrahedrons that share a face (with points from tetrahedrons only sharing vertices but not faces having the ability to alter the exact geometry of the polyhedrons at those edges [22]).

(2) *Robust, piece-wise construction of complex architectures:* because each simplex is transformed into the same class of common auxetic microstructures, by virtue of the construction, the edges and faces of the overlapping segments are co-linear or co-planar across the metamaterial cell edges or faces: this allows a seamless, piece-wise assembly of those triangular or tetrahedral cells into higher-order interconnected metamaterial structures, without exponential increases in the complexity of generating those structures as they grow in number of cells. For example, in the 3D auxetics generated by "shrink-and-wrap", all the open regions between counter-rotating shapes are connected in auxetics generated from a single, connected tetrahedral mesh (Figures 1, 3, and 4), but these properties are retained even if described locally at the level of the individual tetrahedral auxetic cells, without the need of a unified global descriptor or calculation of curvature or connectivity across the entire structure, just at the simplex level.

(3) *Simplified geometric descriptions and analysis:* because of the properties of using the simplex incenter (primarily that they will always project onto each edge or face within the simplex itself at well-defined positions) the geometric description of each microstructure – including positions of coordinates of intersections, range of mobility or bending prior to jamming and internal jam angles (Figures 2J, S2, and S3), and porosities – can all be calculated locally, in terms of each simplex. These can all be readily calculated parametrically from those local points, in terms of edge lengths / face areas, or in terms of internal angles (i.e., Figures S2, and S3). In fact, we find that our local calculations based for range of rotational motion and jamming within the local auxetic cells converges analytically with those global (mechanism- or network-) based approaches for auxetic structures [7], further confirming this construction guarantees a negative Poisson ratio of -1 in a manner that is invariant to the precise geometries of the individual elements.

(4) *Structural motifs in complex geometries:* what becomes apparent after generating (and 3D printing) multiple complex counterrotating metamaterial structures (i.e., Figures 1H, 3A, 4E, and 4F) constructed via the "shrink-and-wrap" approach is that, like the edges of the triangular microstructures come together to form larger counter-rotating polygons, the corners of tetrahedral unit cells come together to form "node"-like polyhedrons with spoke extending into the "shrunken" tetrahedra. Their auxetic properties could be thought of as having arisen from the constrained polyhedral nodes twisting in a manner that pulls the "shrunken" tetrahedra in a way that rotates and tilts them into (or simultaneously away from) the polyhedral cores.

(5) *A natural formulation for piece-wise smoothing of joints and points of contacts between counterrotating shapes:* as designed in this construction (also in others [8]), the counter-rotating shapes come together at a single point of interaction. While this can be approximated using hinges in 2D and 2.5D structures (Figure S1), in practice it is often more useful and necessary in 3D structures for the constructions to generate monolithic metamaterial structures (Figures 1G and 2Fiii). This requires shapes coming together at smoothed / filleted joints instead of hinges. A smoothing of joints in 2D and 3D naturally arises in this construction by introducing quadratic Bézier curves with the "child" vertices of the shrunken triangles/tetrahedra as "control" points (Figure 1F and 2Fii) [23, 24]. At each of those "child" vertices, they can all be used as the coordinates for a Bézier surface connecting shrunken tetrahedra to the polyhedral nodes but localized within each cell for piece-wise generation. We find this approach, introducing Bézier curves and surfaces within the metamaterial cells, works well for three reasons inherent in the "shrink-and-wrap" construction: because the internal geometries within each microstructural cell generated via "shrink-and-wrap" can be simply, locally, and parametrically defined in terms of the coordinates of nearby points and scaling factor λ (i.e., Figure 2G-H) for parametrically-defined Bézier curves; because the edges and faces of the overlapping segments are co-linear or co-planar across, which would be satisfied by quadratic Bézier curves having end-points at those boundaries; and because of the natural symmetries existing within the cells (Figure S2A). Again, this means this smoothing can occur within-cell and piecewise, dramatically simplifying the generation of even extremely complex, smoothed structures (like in Figures 3 and 4) that can then be connected into a single mesh afterwards (see Methods).

(6) *Addressable and programmable mechanical properties across 3 dimensions:* Depending on the degree of smoothness (e.g., Figure 2Fii, characterized by parameter μ; see Methods), the stiffness at each joint connecting the larger polygon/polyhedral nodes to the shrunken triangles/tetrahedra can be tuned with varying joint strength and effective torsional spring constants throughout the construct. These stiffnesses can also be individually tuned throughout the structure: this is a result of the piece-wise nature of the construction and also a result of symmetries and use of each of quadratic Bézier curves terminating at common points within each cell, each of the 3 or 4 joints, respectively, within the same cell can be tuned with different stiffnesses (i.e., 2Fii shows an individual joint being adjusted, orthogonally from the others). This ability to easily and individually control the mechanical stiffness of those joints throughout an entire monolithic mechanical metamaterial can provide a powerful framework for engineering bespoke force distributions, internal rearrangements, mechanical mechanism, or more complex (time- or rate-dependent) behaviors into these structures [1, 3, 9, 26, 27].

To illustrate this construction, its advantages, and its broad applicability in generating 3D mechanical metamaterials, we first generated (and 3D-printed) smoothed 3D mechanical metamaterial structures derived from points describing a regular convex icosahedron [28] containing additional point at the origin (Figure 1B-H) as well as a simple structure defined by two inverted tetrahedra sharing a vertex, rotated 60° relative to one another (Figure 3A). The tetrahedral meshes for these structures derived from those points were simply generated using Delaunay tetrahedralization algorithms [14, 17]. 3D printing and application of forces to compress the structures demonstrated the materials themselves were auxetic (i.e. Figure 1H), that is, contraction in the traverse direction as applied forces; note that this auxetic behavior was also observed in 3D printed 2.5D auxetic structures designed using "shrink-and-wrap" on a constrained Delaunay triangulation to form a non-convex structure (Figure S1). This showed, as expected, the conditions of the piecewise construction are sufficient to impose this negative Poisson ratio property in the counter-rotating metamaterial. In the case of these relatively simple 3D structures, containing essentially a single interior node (the polyhedral generated during the "wrap" step), during contraction it can be seen that the polyhedral nodes rotate to pull together the constrained, counter-rotating "shrunken" tetrahedra in a coordinated manner to collapse the structure in all direction.

For completeness, and in analogy with other works on auxetic metamaterials, we also generated an auxetic "cube" derived from a grid array (3 x 3 x 3) set of points [2, 3, 7] (Figure 3B) that can be readily easily expanded to larger arrays; as well as a high-aspect ratio structure (a Boerdijk–Coxeter helix [29], consisting of 30 stacked regular tetrahedra into a helical column; Figure 3C). To demonstrate the “scale invariance” of the approach, we also generated a “fractal-like” structure, where the coordinates of the bases points of the structure were generated iteratively (Figure 3D): starting with a regular tetrahedron, the point of the insphere center is inserted followed by a Delaunay tetrahedralization; then for each new tetrahedron, the insphere center was then added to the set of generating points. After 3 iterations of this procedure, a fractal-like structure is generated (compare, as well, with the 2D auxetic case in Figure S4).

While those structures were generated from simple, regular, or algorithmically generated sets of points, where we could easily define an interior tetrahedralization explicitly (Boerdijk–Coxeter helix) or using a Delaunay tetrahedralization, many 3D geometric structures used for computational graphics, physical modeling, or additive manufacturing or those generated from 3D scanning or CSG might be represented as triangular shells or meshes with no interior points. To evaluate the performance of the “shrink-and-wrap” approach using tetrahedral meshes of various and increasing complexities, we generated a set of parametrically-defined 3D objects using CSG software (OpenSCAD) including a cylinder (Figure 4A), a sphere (Figure 4B), and a toy model car (Figure 4C; https://en.wikibooks.org/wiki/OpenSCAD_Tutorial/Chapter_2), where first the exterior geometry was defined and output as a triangular surface mesh (containing the list of surface points and triangulations, for example). We also used the surface coordinates of 3D scanned objects (like the “Stanford Bunny” [30]; Figure 4D) containing hundreds or thousands of surface coordinates. Interior tetrahedralizations for those objects where generated from their surface points using common mesh generators for finite element analysis (FEA) or the Fast Tetrahedralization in the Wild (fTetWild) algorithm [12], then the “shrink-and-wrap” algorithm could be immediately applied to these more complex structures. To look more closely at the internal structures of these models, the “bunny” was computationally “sliced” (Figure 4E) to reveal the structural motifs of “nodes” from polyhedral connecting “spokes” of the shrunken tetrahedra integrated throughout its structure, as well as a network of pores interconnected through the entire structure. We were then able to 3D print this mechanical metamaterial “Stanford bunny” using flexible resin for stereolithographic printers (Figure 4F), showing that these designs are robust with structural integrity despite the high levels of structural complexity (Figure 4F right, showing details in close-up) of the overall. Note that the generation of such structures “by hand” or using manual design principles not suitable for automation would be extremely cumbersome, but even in this algorithmic construction the generation of the mechanical metamaterial takes a few minutes on a desktop computer using commonly available and open-source software.

Finally, recent advances in 3D capture, including techniques like Gaussian splatting [15], mean that the 3D coordinates for the surfaces of objects found in the real world can be generated from ~1 minute of video, filmed for example, using a mobile phone camera. However, the objects generated from these types of methods might result in “triangle soup” structures with imperfections or defects that might otherwise limit their ability to be converted to mechanical metamaterial structures [12, 13]. To demonstrate the robustness of our approach, a plush “wolf” doll (wearing a Stony Brook University Department of Biomedical Engineering shirt; Figure 4G) approximately 10 cm in height, and the adult human-sized “Wolfie” state of the athletic mascot of Stony Brook University found on its campus on Long Island, NY, US (Figure 4H) were both filmed using mobile phone camera; converted to 3D surface meshes with Gaussian splatting software; converted to tetrahedral meshes with full internal structures using fTetWild [12]; then transformed into counter-rotating mechanical metamaterial structures in a seamless computational pipeline (Figures 4G-H). These demonstrations show that this robust approach can be used to make potentially any object or geometric structure designed or found in the real world transformable into a monolithic mechanical metamaterial.

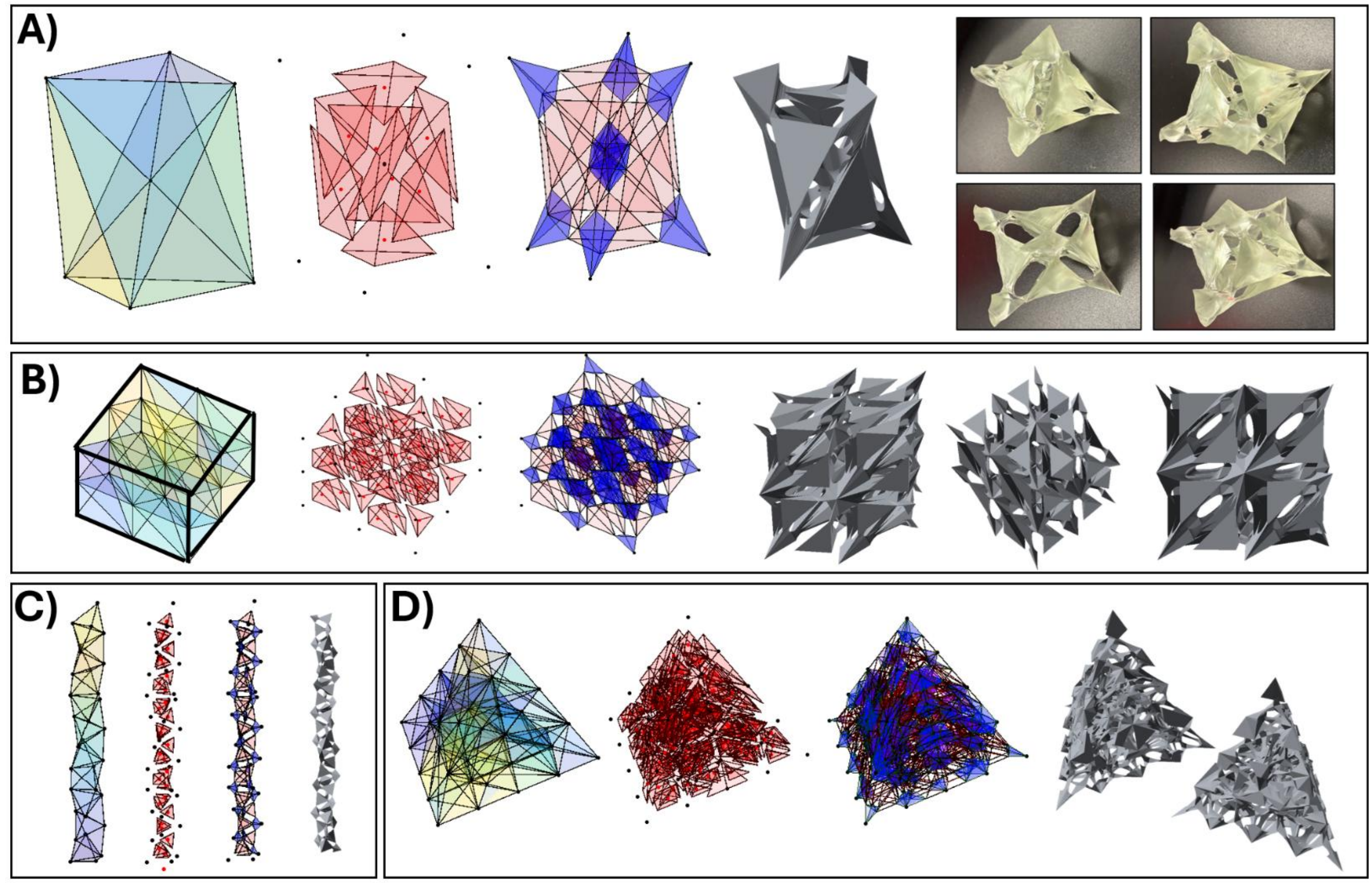

**Figure 3. 3D counter-rotating / auxetic metamaterials derived from relatively simple 3D dimensional shapes.** For each image, from left to right follows the description of the original points / tetrahedral mesh; "shrink" step showing shrunken tetrahedrons (red) towards insphere centers; the "wrap" step of the convex hull formed from the "parent" point and its associated "child" points after the shrinking step (blue); and smoothed 3D auxetic metamaterial (gray). A) Structure defined by 2 inverted tetrahedra sharing a vertex, with base triangles rotated 60° relative to one another. Right) Images of 3D printed auxetic metamaterials, rotated to observe various angles of the exterior and interior. B) A 3 x 3 x 3 cube. C) A Boerdijk–Coxeter helix, which are stacked regular tetrahedra (here 30 tetrahedrons long). D) A "fractal"-like structure from a regular tetrahedron, where the insphere center is inserted into each tetrahedron after Delaunay tetrahedralization, then the Delaunay tetrahedralization performed again on the entire set of points. Structure is after 3 iterations of this procedure.

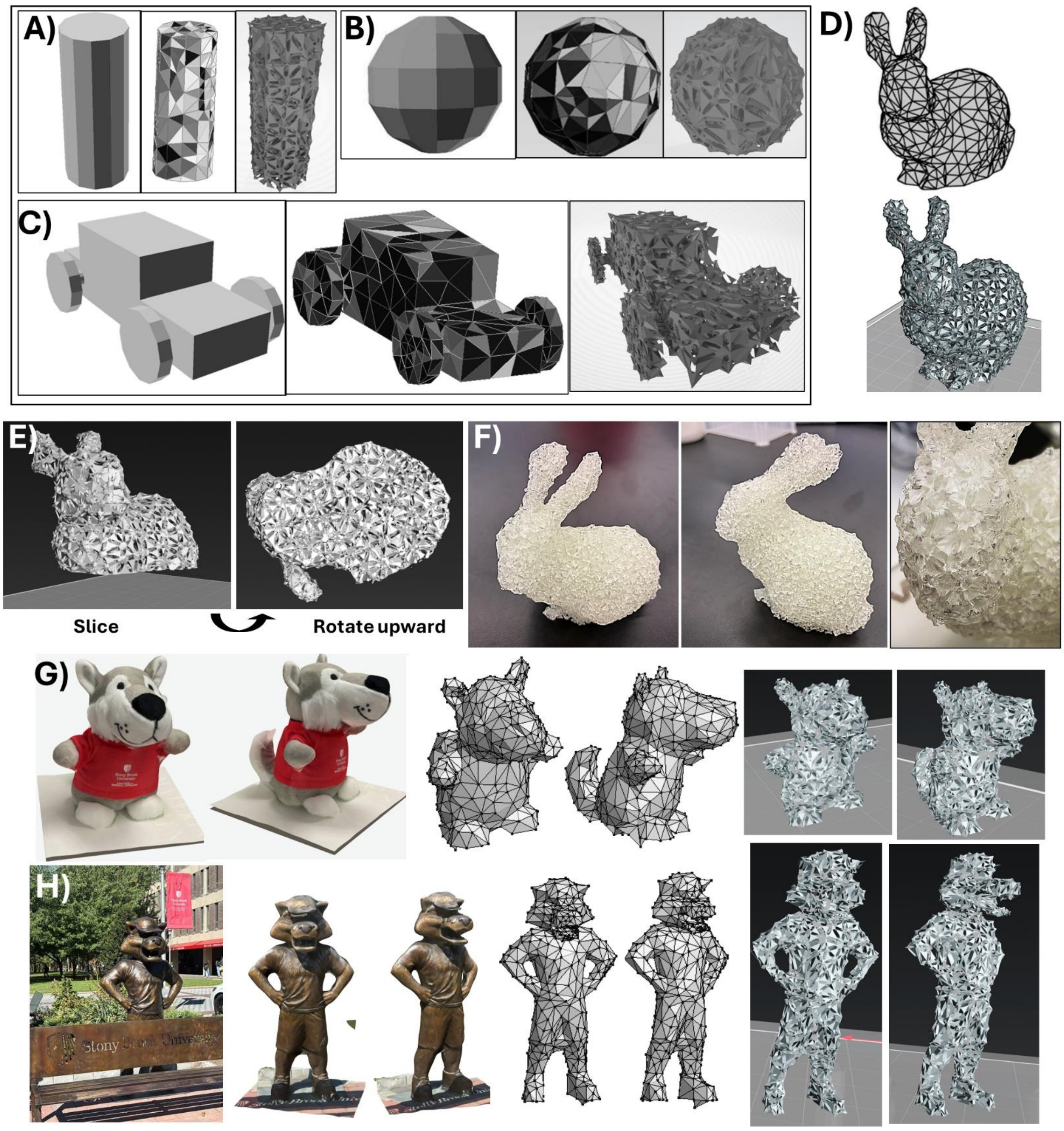

**Figure 4. Smoothed 3D counter-rotating / auxetic metamaterials from complex geometries using piecewise “shrink-and-wrap” construction.** A-C) 3D geometries generated via constructive solid geometry. Geometries were defined using OpenSCAD, exported as triangular surface meshes, then 3D tetrahedralization generated using fTetWild algorithm prior to conversion to auxetic metamaterials using “shrink-and-wrap” construction. Left to write shows structure, surface mesh (after tetrahedralization) then auxetic metamaterial after generation. A) A 10 x 2 cylinder. B) A sphere. C) A rudimentary car (from OpenSCAD Tutorial/Chapter 2). D) The Stanford Bunny (with ~10% of surface points) surface mesh (above) and 3D auxetic Stanford Bunny (below). E) The 3D auxetic model derived from the Stanford Bunny model sliced and rotated to show fully auxetic structure throughout the interior of the model. F) 3D-printed Stanford Bunny auxetic metamaterial (right) close-up showing fine features of shrunken tetrahedra and connecting polyhedral “nodes”. G-H) 3D auxetic metamaterials

generated from models created using videos captured by mobile phone camera converted to 3D structures. G) Right two) A stuffed wolf doll, wearing a Stony Brook University Department of Biomedical Engineering shirt, (middle two) Surface mesh generated after conversion to digital 3D structure. (right two) 3D auxetic mechanical metamaterial model. H) (left) “Wolfie” statue on Stony Brook University campus on Long Island, NY, USA. (2 and 3 from left) 3D imaged structures after video capture and conversion. (4 and 5 from left) Surface mesh generated after conversion to digital 3D structure. (right two) 3D auxetic mechanical metamaterial model.

## Conclusions:

With the imposition of a simple and common design constraint (that the structure be decomposable into triangular or tetrahedral meshes [12, 14, 17]), here we show a straightforward method to convert the components of any triangular or tetrahedral meshes into a member of a geometrically-diverse family of microstructural cells, transforming any such structure into a mechanical metamaterial of counter-rotating polygons or polyhedrons. The piecewise and locally-determined nature of the metamaterial cells generated through "shrink-and-wrap" approach provides a powerful basis for the analysis, design, and systematic study of these structures in 3D without snowballing complexity (e.g., Figures S2 - 4). For example, while the overall auxetic nature of the structures remains independent of the precise internal geometry by virtue of its construction [7], we might expect local density or angles between might affect the precise mechanics and properties of the resulting mechanism in terms of jamming (Figure S2) [8], porosity (Figures S3 and S4), and how they respond to more complex dynamics [1-3, 5, 6, 9, 10, 31-33]. The ability to easily and individually control the mechanical stiffness of specific joints naturally built into the constructure can also provide a powerful framework to integrate more sophisticated mechanical mechanisms into their monolithic structures [26, 27]. Because of the resulting simplexes can be continuously split by adding in new points (e.g., Figures 2D and S4 of the "fractal"-like structures), we expect that this construction approach can lend itself well to fine-tuning force responses and realizing a myriad of desired mechanics with increasing levels of spatial approximations, by piggy-backing off known tools for generating meshes at different required spatial resolutions; additionally, because the transformation is algorithmic and the global structure can be simply determined by the set of points (and their connectivity in a mesh), this reduction in the design space can likely be readily adapted to other optimization techniques and AI-based design approaches [6, 9-11]. Hence, we expect this approach can significantly advance the application, engineering, and deployment of mechanical metamaterials for a variety of their emerging applications.

## Acknowledgements.

We wish to acknowledge the current and former members of the Stony Brook University Vertically Integrated Projects (VIP) program Mechanical Metamaterials and Kinetic Devices team (with authors F. B., I. V., and N. K. also as members and corresponding author E. A. J. as team mentor), for helpful discussions, feedback, collaboration, and/or assistance in preparation of this manuscript. We also wish to acknowledge Sophy Meija (University of North Carolina at Greensboro) for 3D printing early prototypes of these metamaterials and Prof. Jeffery Alston (North Carolina A&T State University) for helpful discussions. We also wish to acknowledge members of the Josephs laboratory for helpful discussions and feedback, as well as L. M. J. and A. S. J. for helpful discussion and donation of materials to 3D scan. Funds for this research were provided by the National Institute of General Medical Sciences (R35GM133483 to E. A. J.), funds from the SUNY Empire Innovation Program and the Department of Biomedical Engineering at Stony Brook University, the VIP program at Stony Brook University, and the Undergraduate Research & Creative Activities (URECA) program at Stony Brook University. (to N. K.).

## References.

**Supplementary information.**

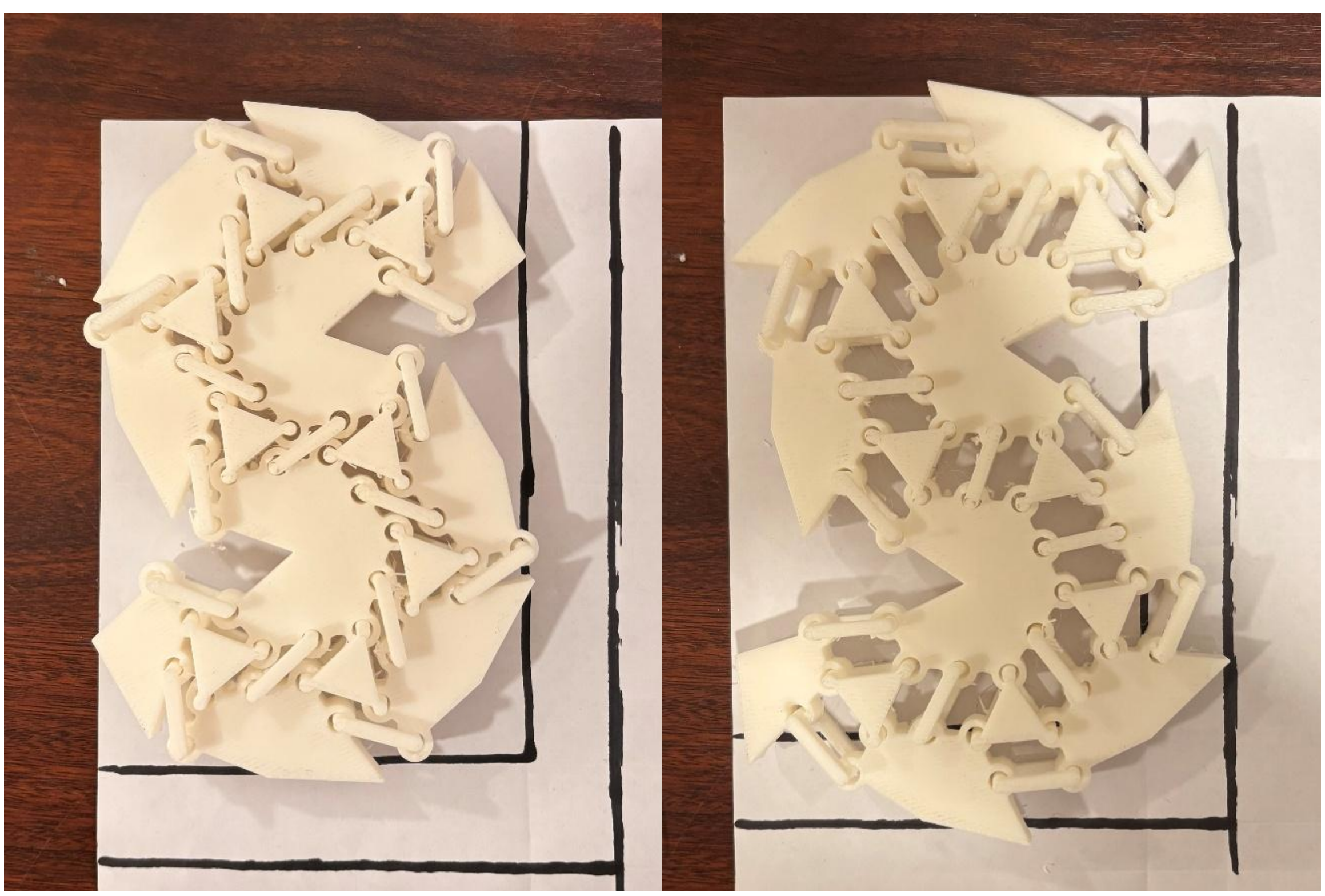

**Figure S1. 3D printed 2.5D auxetic structures designed using "shrink-and-wrap" on a constrained Delaunay triangulation to form a non-convex structure.** Left) Compressed, and right) expanded) demonstrating, as expected the conditions of the piecewise construction are sufficient to impose this negative Poisson ratio (auxetic) property caused by counter-rotating shapes. The counter-rotating polygons come together using hinges, although in practice it is often more useful for the constructions to generate monolithic metamaterial structures (i.e. Figure 2Fiii).

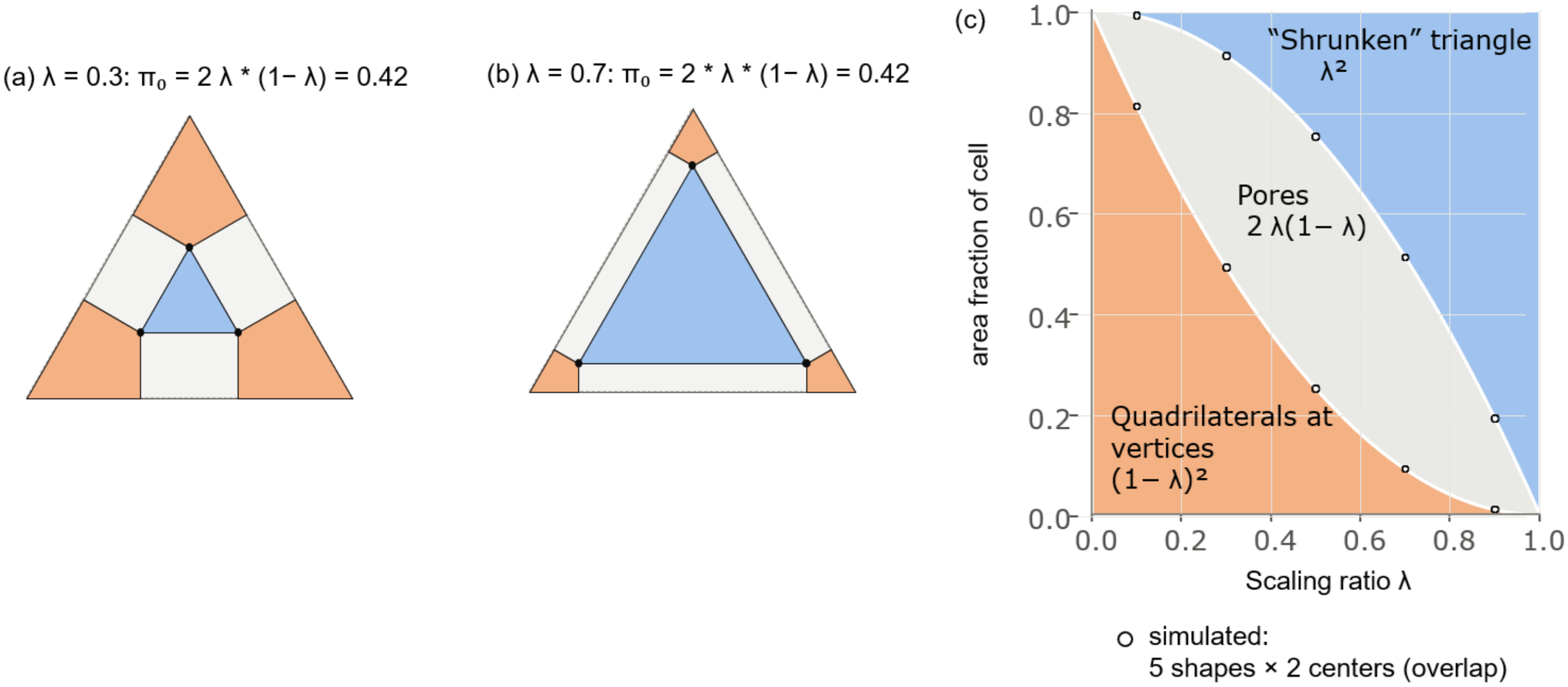

**Figure S2. Close-up of the family of counter-rotating auxetic microstructural cells in 2D and 2.5D.** A-B) Effect of changing the scaling factor λ from A) 0.3 (smaller "shrunken" triangle) to B) 0.7 (larger "shrunken" triangle). Above, the fraction of empty space / the "porosity" ($\pi$; with $\pi_0$ referring to porosity at 0° rotation angle between "shrunken" triangle and the counter-rotating quadrilaterals at vertices) obeys a simple relation that can be derived, dependent only on λ. While the example here is equilateral, this relationship holds regardless of the precise geometry of the triangular cell (that is, regardless of whether the triangles are isosceles, acute, obtuse, right, and scalene). C) Fraction of the auxetic cell makeup by components ("shrunken" triangle, quadrilaterals at vertices, or empty space / pores) as a function of λ.

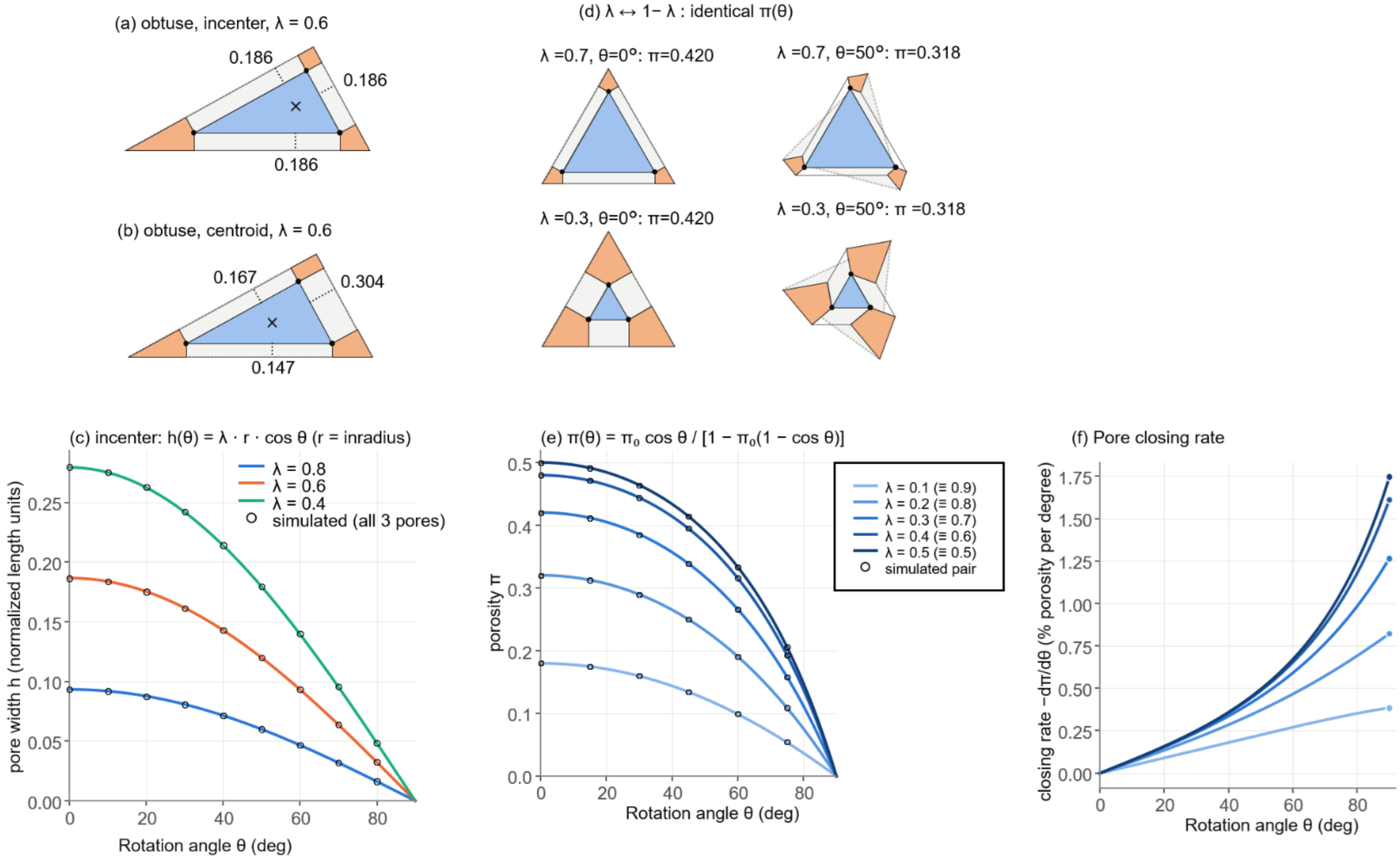

**Figure S3. Calculations of geometries and changes in porosity with counter-rotation, from the vantage of individual 2D auxetic cells.** A - B) An example of one of the analytical / geometric advantages of using the triangle incenter (A) as opposed to (B) triangle centroid in constructing the auxetic cell is that the pore widths (length of the empty space from triangle edge to edge of the "shrunken" triangles are the same for all three sides, introducing a level of symmetry when incenter is used. That is not the case for when the centroid (or excenter) are used in the construction. C-D) Pore widths compress as the polygons are rotated relative to one another, closing the pores/empty space to result in the collapsed auxetic structures. Derived expressions of these terms match simulated values. D-E) Regardless of the precise dimensions and geometries of the triangles (i.e., isosceles, acute, obtuse, right, and scalene), changes in porosities π ($\pi_0$ referring to porosity at 0° rotation angle between "shrunken" triangle and the counter-rotating quadrilaterals at vertices) (D) and the rate of closure (E) obey a common analytical expression that can be derived from the individual auxetic cell level.

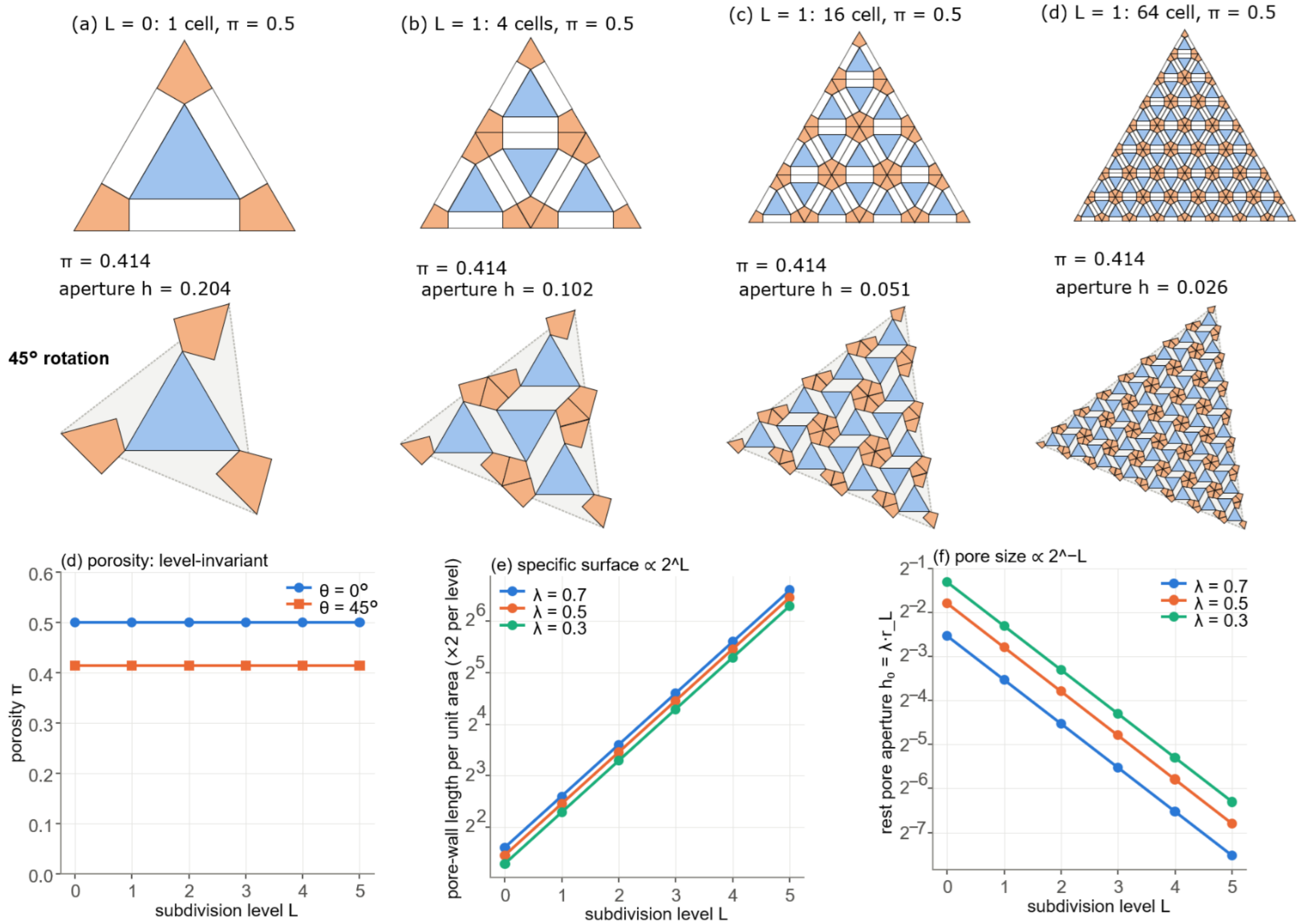

**Figure S4. Fractal triangular auxetic structures -- compare with Figure 2D for 3D auxetics.** A-D) Increasing fractal levels (L) from 0 (A), 1 (B), 2 (C), and 3 (D) by iterated splitting of the triangular cells into 4 unit cells by connecting midpoints. (Above) No rotation, (Below) with 45° rotation. D) Increasing the fractal level does not change the overall porosity of the (initial) cells, even though it significantly does (E) increase surface area and (F) decrease pore size. Compare, for example with the “fractal” generation of 3D auxetics by iterative insertion of incenter points into each tetrahedron, followed by re-tetrahedralization during “shrink-and-wrap”.